\documentclass[preprintnumbers,amsmath,11pt,amssymb,floatfix,superscriptaddress,nofootinbib]{article}

\def\mytitle#1{\setcounter{equation}{0}
\setcounter{footnote}{0}
\begin{flushleft}\Large\textbf{#1}\end{flushleft}
\vspace{0.25cm}}
\def\myname#1{\leftline{{\large #1}}\vspace{-0.13cm}}
\def\myplace#1#2{\small\begin{flushleft}\textit{#1}\\
\texttt{#2}\end{flushleft}}

\def\myclassification#1{\small\noindent
Keywords :
       #1\vspace{0.5cm}}
\usepackage{graphicx}
\begin{document}

\mytitle{Black Hole Thermodynamic Products in Einstein Gauss Bonnet Gravity}

\myname{$Abhijit~ Mandal^{*}$\footnote{abhijitmandal.math@gmail.com}, $Ritabrata~
Biswas^{\dag}$\footnote{biswas.ritabrata@gmail.com}}
\myplace{*Department of Mathematics, Jadavpur University, Kolkata $-$ 700 032, India.\\$\dag$ Department of Mathematics, Bankura University, Bankura $-$ 722 146, India.}{} 
 
\begin{abstract}
By now, there are many hints from string theory that collective excitations of solitonic objects can be described by effective low energy theories. The entropy of general rotating black holes in five dimensions may be interpreted as an indication that, it derives from two independent microscopic contributions and each of these may be attributed to a gas of strings \cite{Cevtic3}. In the present work, we consider a charged black hole in five dimensional Einstein Gauss Bonnet gravity. In spite of presenting the thermodynamic quantities' product as summation/ subtraction of two independent integers, our motive is to check whether the product of the same quantity for event horizon and Cauchy horizon is free of mass, i.e., global, or not. We derive the thermodynamic products of characteristic parameters to mark which are global. We further interpret the stability of the black holes by computing the specific heat for both horizons. Stable and unstable phases of horizons are pointed out. The phase transitions with respect to the charge in nature of specific heat are also observed. All these calculation might be helpful to understand the microscopic nature of such black holes.

\end{abstract}
\myclassification{ Black hole physics; Thermodynamics; Thermodynamic processes, equation of state; Relativity and gravitation. }

\section{Introduction}
For a black hole (BH hereafter) if mass is the only characteristic parameter, we get the event horizon. Crossing through this event horizon shows swapping of the role of temparal and spatial co-ordinates. It's a singular region around the $r=0$ singularity. This will not let any information to come outside from inside.

	 But except the mass whenever we introduce other parameters the scenario changes. Say, for Reissner Nordstr$\ddot{o}$m BH, if the charge of the BH is less than its mass(measured in geometric units $G=c=1$), then two horizons exist. Between the two horizons, space acts like a waterfall, falling faster than the speed of light.

Using Boyer-Lindquist co-ordinates for Kerr metric, we see when $r=0$ and $\theta = \frac{\pi}{2}$ (i.e., $r^2 + a^2 sin^2 \theta =0$) we get singularities. It gives
$$1-\frac{2m}{r} + \frac{a^2}{r^2}=0 \Rightarrow r=r_\pm \equiv m \pm \sqrt{m^2 - a^2}.$$
From outer layer to inner layer we see, one by one, different general relativistic backgrounds :\newline
(i) Outer ergosurface  $ (r^+_E = m + \sqrt{m^2 - a^2 cos^2 \theta})$ \newline
(ii) Ergoregion \newline
(iii) Outer event horizon  $(r_+ = m + \sqrt{m^2 - a^2 })$ \newline
(iv) Inner event horizon  $(r_- = m - \sqrt{m^2 - a^2 })$ \newline
(v) Inner ergossurface  $(r^-_E = m - \sqrt{m^2 - a^2cos^2 \theta })$ \newline

	Now, General Relativity(GR hereafter) admits well posed initial value problem. If we provide some initial data on the spacelike hyper- surface, say $S$, the solution to the Einstein equation is uniquely determined everywhere within the domain of dependence $D(S)$ of $S$. There are several examples of space times for which $D(S)$ is unable to cover the total manifold $M$. This emplies that even if we provide a suitable initial data on $S$, GR is unable to forecast the evolution of spacetime beyond $D(S)$ in these cases. The boundary of $D(S)$ denoted by $H(S)$, is known as Cauchy Horizon(CH hereafter). This horizon is acting as a barrier line of two regions : one, where GR is able to predict the evolution and second, where predictability of the field equations is lost. For a BH we say the second large horizon to be the Cauchy horizon. Infinite proper time compression effect will make the journey from event horizon to CH unstable. The casual past i.e., the set of all source points from which time like curves are coming to CH, contains the entire universe external to the concerned BH. So any observer who is approaching to the CH, few moments before he crosses the CH, sees the entire history of external universe to flash infront of eyes.

   	Besides all the physical understandings of CH, there is a broad mathematical view of it. Since $1990$'s, a consistent picture of a classically stable CH \cite{Chambers} in BH$-$ de Sitter space time has emerged. Both linear studies and non-linear back reaction calculation indicate that the CH is stable. In other reference like \cite{Brady1}, based on the idea of infinite time compression effects, it was suggested that CH will always be stable. The assumption of this work agreed that stability persists for only a finite though non-zero measure of the parameter space $(M, Q, J, \Lambda)$ of the BH, infringing on the very spirit of the strong cosmic censorship hypothesis.
   	
   	Current analysis of BH thermodynamics requires to trust on the universal symmetry arguments that follow from the analysis of $3D$ gravity in AdS space time \cite{Brown}.
   	
   	The microscopic degrees of freedom of the BH are described in terms of those of a conformal field theory - without gravity - living in the boundary. This construction gave the inner $(A_-)$ and outer $(A_+)$ Killing horizons of BHs obeying the product formula 
  $$\frac{A_+ A_-}{(8\pi G_3)^2} = N_{R} - N_{L}~~,$$
where $N_{R}$ and $N_{L}$ are number of right and left moving excitations of the $2D$ conformal field theory. Generalizing for any asymptotically flat BH in $d$ - spacetime dimensions \cite{Larsen}, 
$$\frac{A_+ A_-}{(8\pi G_d)^2} \in Z~~, ~the ~set~of~integers~. $$
In a nut shell, this says the product of areas is independent of BH mass and solely depends on the quantized charges. This suggestive general relation for BHs encouraged purely gravitational investigation of the product of area of BHs \cite{Cvetic1}. For Kerr-Newman BH area - product is been investigated \cite{Ansorg}.

	Contrary to the outer horizon thermodynamics, it is not very clear where the inner - horizon has any relevance for a statistical accounting of the BH entropy. First law for the inner horizon is schematically  
\begin{equation}\label{inner_horizon}
-dM = T_{-}\frac{dA_{-}}{4G_{5}} - (\Omega^{-} dJ + \phi^{-}_{E} dQ + \phi^{-}_{m} dq)~~.
\end{equation}	
	The reference \cite{Curir} has studied the above relation for Kerr BHs and references \cite{Cvetic2} studied the same for more generalized BH. For the sake of comparison the first law for the outer horizon is 
\begin{equation}\label{outer_horizon}
dM = T_{+}\frac{dA_{+}}{4G_{5}} + (\Omega^{+} dJ + \phi^{+}_{E} dQ + \phi^{+}_{m} dq).
\end{equation}	
	The minus signs in (\ref{inner_horizon}) are due to the Killing horizon vector field being spacelike inside the BH event horizon. Keeping similarity to the negative energies within ergosphere, negative energy ($-M$) is assigned to the inner horizon as well. For the perturbatively unstable nature of the inner horizon one may follow \cite{Marlof, Castro1}. 
				If someone considers a number of different higher curvature modifications of the Einstein Hilbert action, the entropy will not generally be proportional to horizon area and the product relations for the horizon areas and entropies will no longer be equivalent. According to \cite{Wald1}, BH entropy can be viewed as a Noether charge, it might be more natural to except that the entropy product formula, rather than the product of the areas would be the correct generalization for modification to Einstein gravity. There exist(s) such references like \cite{Castro2}, where considering the holding of entropy product formula as a `success', a try to find `failures' has been taken, i.e., where the product of the horizon entropy depends on the mass. For such `failures' Smarr relation is modified. Though there is no obvious relation between the horizon entropy products and Smarr relations, but for these failing cases both are drastically modified by the presence of higher curvature terms

Our motivation is to check the thermodynamic products for some higher dimentional modified gravities. Einstein Gauss Bonnet gravity is a modification of the Einstein Hilbert action to include the Gauss Bonnet term 
\begin{equation}
 R_{GB}=R^2 - 4R_{\mu\nu}R^{\mu\nu} + R_{\mu\nu\gamma\delta}R^{\mu\nu\gamma\delta}.
\end{equation} 
 Here $R$, $R _{\mu \nu}$ and $R_{\mu\nu\gamma\delta}$ are, respectively, the Ricci scalar, the Ricci tensor and the Riemann tensor of $M$.
 
 The action of the Einstein GaussBonnet(EGB hereafter) Gravity in $5$ dimensional space time ($M,g_{ij}$) can be written as \cite{Cai} (Taking $8\pi G = c = 1$ as unit)
\begin{equation}
S = \frac{1}{2}\int_{M} d^5 x\sqrt{-g}[R + \alpha R_{GB} + L_{matter}],
\end{equation}
$\alpha$ is the coupling constant of the GB term having dimension $(length)^2 (\alpha \geq 0)$.
 
  $L_{matter} = F_{\mu\nu} F^{\mu\nu}$ is the matter Lagrangian where $F_{\mu\nu} = \delta_{\mu} A_{\nu} - \delta_{\nu} A_{\mu}$ is the electromagnetic tensor field, $A_{\mu}$ is the vector potential. 
	The gravitational and electromagnetic field equations obtained by varying the
action (i.e., $\delta S = 0$) with respect to $g_{\mu\nu}$ and $F_{\mu\nu}$ are
\begin{equation}\label{EGB}
G_{\mu\nu}= T_{\mu\nu}^{EM} + T_{\mu\nu}^{GB}
\end{equation}
and
\begin{equation}\label{Maxweel_equation}
\nabla_{\mu} F_{\nu}^{\mu} = 0,
\end{equation}
where, $T_{\mu\nu} = \alpha H_{\mu\nu}$, where $H_{\mu\nu}$ is the Lovelock tensor given by,
\begin{equation}
H_{\mu\nu} = 2[R R_{\mu\nu} - 2R_{\lambda\nu}R_{\nu}^{\lambda} - 2R^{\gamma\delta}R_{\mu\gamma\nu\delta} + R_{\mu}^{\alpha\beta\gamma}R_{\nu\alpha\beta\gamma}] - \frac{1}{2} g_{\mu\nu}R_{GB}
\end{equation}
and $T_{\mu\nu} = 2F_{\mu}^{\lambda} F_{\lambda\nu} − \frac{1}{2} F_{\lambda\sigma} F^{\lambda\sigma} g_{\mu\nu}$ is the electromagnetic stress tensor.

The Gauss Bonnet Lagrangian $R_{GB}$ is only non-trivial in $(4+1)D$ or greater, and as such, only applies to extra dimensional models. In $(3+1)D$ and lower it reduces to a topological surface term.

We now proceed to solve the field equations (\ref{EGB}) for the five-dimensional static spherically symmetric space time with the line element
\begin{equation}
ds^2 = -f(r) dt^2 + \frac{dr^2}{f(r)} + r^2 d\Omega ^2_3
\end{equation}
where $d\Omega ^2_3$ is the metric of a $3D$ hyper surface with the constant curvature $6K$ having an explicit form
\begin{equation}
d\Omega ^2_3 = 
\left\{
\begin{array}{lll} 
d\theta_1^2 + sin^2 \theta_1 (d\theta_2^2 + sin^2 \theta_2 d\theta_3^2 ) , \mbox{(K= 1)}\\\\
d\theta_1^2 + sin^2 \theta_1 (d\theta_2^2 + sinh^2 \theta_2 d\theta_3^2 ), \mbox{(K= -1)}\\\\
\alpha^{-1} dx^2 + \sum^2_{i=1} d\phi^2_i,~~~~~~~~~ \mbox{(K=0)}
\end{array}
\right.
\end{equation}
where the coordinate $x$ has the dimension of length while the angular coordinates $(\theta_1, \theta_2, \theta_3)$ and $(\phi_1 , \theta_2)$ are dimensionless, with ranges
$\theta_1, \theta_2 : ~~~[0, \pi]~~~~~~~~\theta_3, \phi_1 , \phi_2 : ~~~[0, 2\pi].$

If we assume that there exists a charge $q$ at $r = 0$ (note that $q$ is a point charge for $K = \pm 1$ and is the charge density of a line charge for $K = 0$), then the vector potential may be chosen to be
$A_\mu = \phi(r)\delta_\mu^0.$
Now using the Maxwell equation (\ref{Maxweel_equation}), the differential equation for $\phi(r)$ becomes
\begin{equation}
r\frac{d^2\phi}{dr^2} + 3\frac{d\phi}{dr} = 0 \Rightarrow \phi(r) = - \frac{q}{2r^2},
\end{equation}
where the Gauss law has been used to determine the integration constant. Now the
metric function for EGB gravity, say $f_{EGB} (r)$ can be obtained by solving the field equations (\ref{EGB}) as \cite{Dehghani}
\begin{equation}\label{BH_solution}
f_{EGB} (r) = K +\frac{r^2}{4\alpha}\left[1 \pm \sqrt{1+ \frac{8 \alpha(M +2\alpha|K|)}{r^4}-\frac{8\alpha q^2}{3r^6}}\right].
\end{equation}
For the black hole solution (\ref{BH_solution}) in GB theory, the horizons correspond to $f_{EGB} (r)$ and we obtain
\begin{equation}
r_{h/c}= \frac{1}{2}\left[\sqrt{M+\frac{2q}{\sqrt{3}}} \pm \sqrt{M-\frac{2q}{\sqrt{3}}} \right],
\end{equation}
where this $h$ sign corresponds to event horizon $(r_h)$ and the $c$ sign corresponds to Cauchy horizon $(r_c)$.This is to be followed that the radii of horizons are independent of the coupling parameter $\alpha$.

EGB gravity is the most simple non-trivial example of Lovelock gravity. We want to investigate the thermodynamic product of a BH lying in EGB gravity. 
In the next parts of this letter we will calculate different thermodynamic products and analyse the stability of such BHs. At the end we will conclude what we have done.
\section{Thermodynamic Product Analysis for EGB BH with classical Area-Entropy Relation} 
Product of the radii of the event horizon and CH is
\begin{equation}
r_h r_c = \frac{q}{\sqrt{3}}, 
\end{equation}
which is a BH mass free global quantity.
The surface area of the event horizon is given by
\begin{equation}
A_{h/c}=r^3_{h/c}\int_{\theta=0}^{\pi} \int_{\phi=0}^{\pi}\int_{\varphi=0}^{2\pi} sin^{2} \theta~ sin\phi ~d\theta ~d\phi ~d\varphi = 2 \pi^2 r^3_{h/c}.
\end{equation}
So,
$$A_h A_c = 4 \pi^2 r^3_{h}r^3_{c} = \frac{4}{3\sqrt{3}} \pi^4 q^3.$$
Again this is another mass free global quantity.
Now, the entropy will take the form \cite{Falcke}
\begin{equation}
S_{h/c} = \frac{K_B A_{h/c}}{4 G \hbar} = \frac{K_B}{2 G \hbar}  \pi^2 r^3_{h/c}.
\end{equation}
Now, setting our units such a way that we have $\frac{K_B\pi^2}{2 G \hbar} =1$,entropy becomes
\begin{equation} \label{entropy_product}
S_{h/c}=r^3_{h/c}=\frac{1}{8}\left[\sqrt{M+\frac{2q}{\sqrt{3}}} \pm \sqrt{M-\frac{2q}{\sqrt{3}}} \right]^3~~~\Rightarrow~~ S_hS_c = r_h^3 r_c^3 = \frac{q^3}{3\sqrt{3}}.
\end{equation}

The temperature of the BH is given by
\begin{equation}
T_{h/c}=\frac{\partial M}{\partial S_{h/c}} = \frac{1}{\frac{\partial S_{h/c}}{\partial r_{h/c}}\frac{\partial r_{h/c}}{\partial M}} = \pm \frac{4}{3} \frac{\sqrt{M^2 - \frac{4 q^2}{3}}}{r_{h/c}^3}~~~~~and~we~have~,~
\end{equation}
\begin{equation}
T_hT_c = - \frac{16}{\sqrt{3}q^3} \left(M^2 - \frac{4 q^2}{3}\right).
\end{equation}
So this is depending on mass of BH and hence is not a global property.
The free energy will take the form
\begin{equation}
F_{h/c}=M-T_{h/c}S_{h/c}=M \mp \frac{4}{3} \sqrt{M^2 - \frac{4q^2}{3}}~~~ \Rightarrow~~ F_hF_c = \frac{1}{27}( 64q^2 - 21M^2 ).
\end{equation}
This, like temperature is not a global property.
Now proceeding towards heat capacity, we find 
\begin{equation}
C_{h/c}=T_{h/c}\frac{\partial S_{h/c}}{\partial T_{h/c}}=\frac{\partial M}{\partial S_{h/c}}\frac{\partial S_{h/c}}{\partial T_{h/c}}=\frac{\partial M}{\partial T_{h/c}}=\frac{1}{\frac{\partial T_{h/c}}{\partial M}}
= \frac{3 r^3_{h/c} \sqrt{M^2 - \frac{4 q^2}{3}}}{16\left(\pm 2M - 3 \sqrt{M^2 - \frac{4q^2}{3}}\right)}.
\end{equation}
So the product is 
$$C_hC_c = \frac{\sqrt{3} q^3 \left( M^2 - \frac{4 q^2}{3}\right)}{256 \left(5 M^2 - 12 q^2\right)},$$
which is depending upon the mass $M$ of the BH again and is not a global property.

\begin{figure}[h]
\begin{center}

~~~~~~~~~~Fig.1a~~~~~~~~~~~~~~~~~~~~~~~~~~~Fig.1b~~~~~\\
\includegraphics[height=1.5in, width=2.2in]{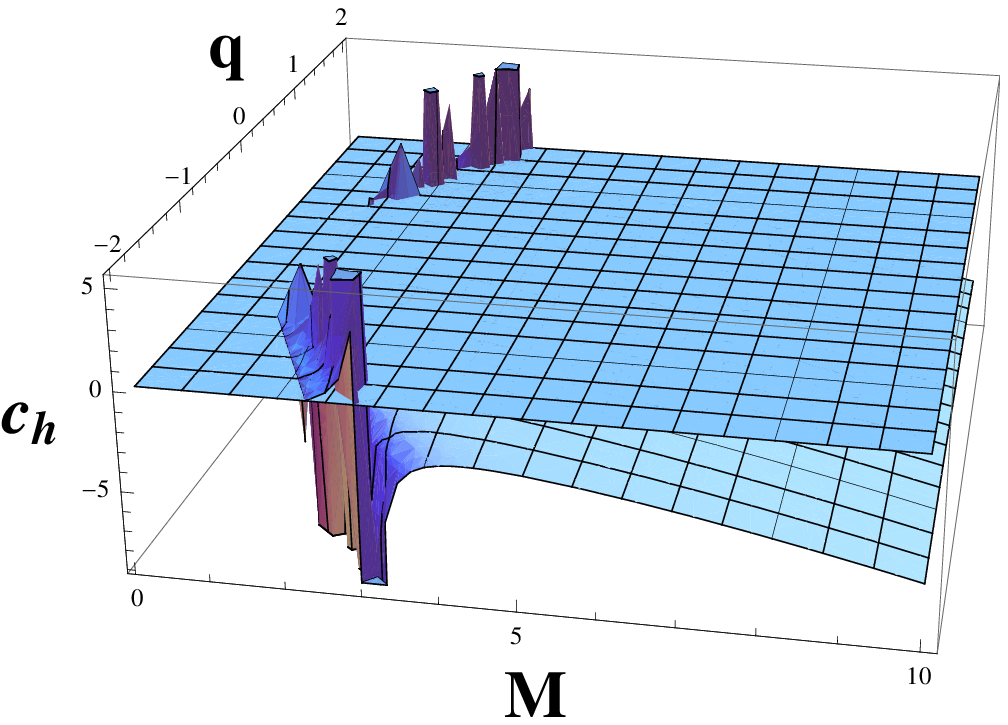}~~~~
\vspace{.1cm}
\includegraphics[height=1.5in, width=2.2in]{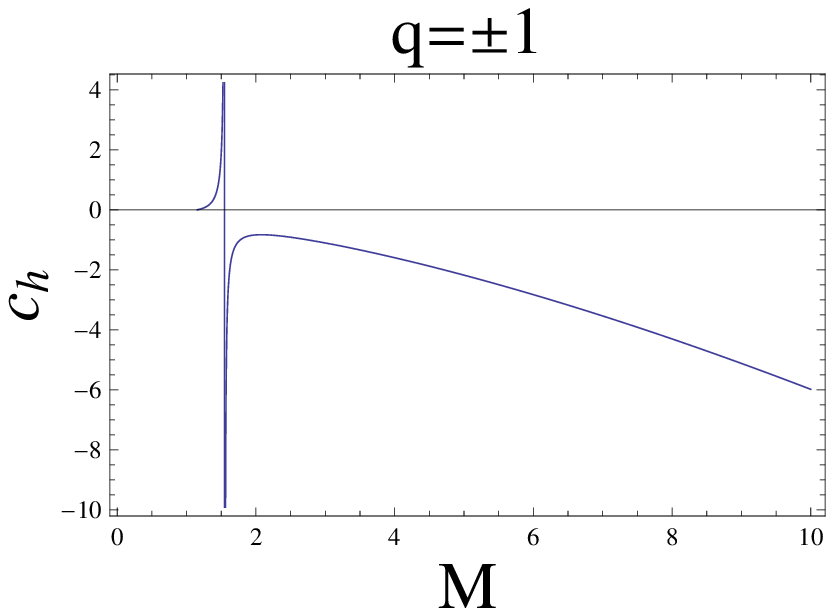}~~\\
Fig $1a$ represent the variation of specific heat($C_h$) for event horizon with respect to $M$ and $q$.\\
Fig $1b$ represent the variation of specific heat($C_h$) for event horizon with respect to $M$ for charge $q=\pm 1$.

\end{center}
\end{figure}

In Fig. $1a$, we observe that for small charge and small mass of BHs we get $C_h$ negative i.e., unstable BHs. Now keeping $q$ constant if we increase $M$, BHs remain unstable. But increment of $|q|$ gives us a stable at very begining which after passing through a second order phase transition become unstable.

Now we will analyse Fig. $1b$ where the specific heat for event horizon($C_h$) is plotted with respect to the mass($M$) for charge $q = \pm 1$. In Fig. $1b$ when the mass, $M$, is small, the specific heat for event horizon is positive but if we increase the $M$, $C_h$ become negative through a second order phase transition and remains same for large value of $M$ i.e., initially for small value of $M$, BHs are stable but if we increase tha value of $M$ BHs become unstable through a second order phase transition and remains unstable.

\begin{figure}[h]
\begin{center}

~~~~~~~Fig.2a~~~~~~~~~~~~~~~~~~~~~~~~~~~~Fig.2b~~~~~~~~~~~~~~~~~~~~~~~~~~~Fig.2c~\\
\includegraphics[height=1.5in, width=2.2in]{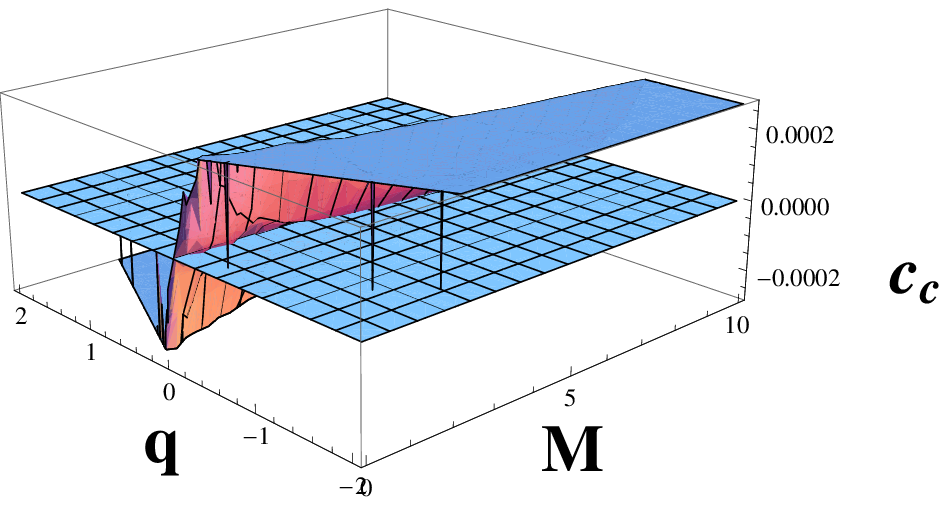}~~~
\vspace{.1cm}
\includegraphics[height=1.5in, width=2in]{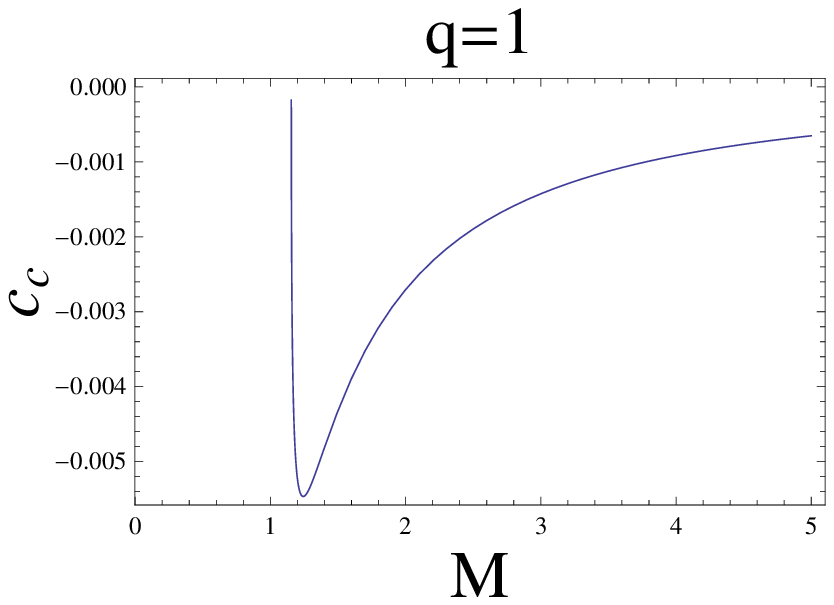}~~~
\vspace{.1cm}
\includegraphics[height=1.5in, width=2in]{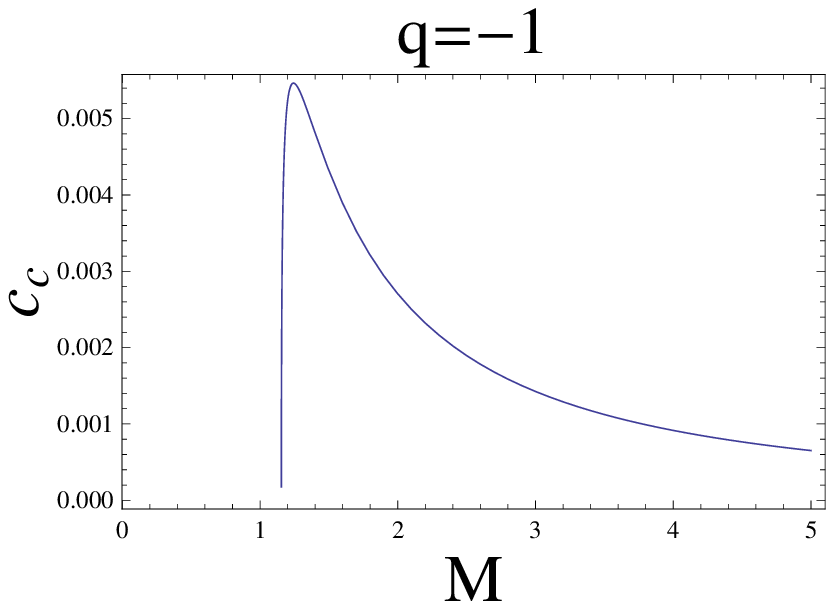}~~\\
Fig. $2a$ represent the variation of specific heat($C_c$) for CH with respect to $M$ and $q$.\\
Fig. $2b - c$ represent the variation of specific heat($C_c$) for CH with respect to mass $M$ for charge $q = 1, -1$ respectively .

\end{center}
\end{figure}

Now we will analyse Fig. $2a$ where the specific heat($C_c$) for CH is plotted with respect to the $M$ and here the variation with respect to the charge $q$ has also been taken. We should notice in Fig. $2a$ the fact that for positive charge $C_c$ is negative and for negative charge this is positive.

Here in Fig. $2b$ the general trend of $C_c$ vs $M$ curve for $q=1$ is primarily strictly decreasing and slowly increasing latter. If we increase $M$ gradually, $C_c$ is a decreasing function at first then after reaching a local minima it increases. Here overall the sign of the $C_c$ is negative.

According to Fig. $2c$, for $q= -1$, the general trend of $C_c$ vs $M$ is primarily increasing and slowly decreasing latter. In Fig. $2c$ the value of $C_c$ remains always negative, i.e., the part of BH surrounded by CH is always unstable. Here if we increase $M$ gradually, $C_c$ is a increasing function at first then after reaching a local maxima it decreases. If $M=\pm \frac{2q}{\sqrt{3}}$, even temperature and $C_c$ for both event horizon and CH will become independent of mass.
\section{Discussion}
In existing literatures \cite{Biswas}, it was already found that Modified gravity BHs are thermodynamically unstable on a long run specially when they are large. The same result is established here. But the space time confined within  event horizon is stable when radius of event horizon is small. Latter a second order phase transition makes them unstable. However, as we stated this is not a new statement in the study of Modified gravity BH thermodynamics. But if we follow the space time confined inside CH, this is unstable if BH is positively charged and there is a possibility to have ever stable part confined in CH if this is negatively charged. This case is not symmetric with $q$(we mean symmetric with $q=0$ plane). Rather these are $180^\circ$ rotational image of each other. This some how says positively charged part inside CH is turbulent more than the negatively charged one. About the thermodynamic product part, the entropy(i.e., the area) of EGB BHs are global.

 It is a characteristic property of string models that the entropy is the sum of contributions from left and right moving excitations of the string. The BH geometry exhibits an analogous structure. Standard thermodynamic variables which are defined at the outer event horizons, are mirrored by an independent set of thermodynamic variables, defined at the inner event horizon\cite{Horowitz}. Again from (\ref{entropy_product}) it is to be followed for integral/ rational values of $q$, $S_h \times S_c$ can not be represented as sum/ subtraction of two integers. If we were able to do so, we could have say that the terms, sum of which is $S_h \times S_c$, are contributions from left and right moving modes. This again could have been treated  as a notification towards a weak interaction of these two kind of modes. As we are considering five dimensions, our entropy is the third order of the radii. This makes it irrational. Here we can say that the interaction is not weak.

\vspace{1 in}
{\bf Acknowledgement:}
RB thanks Inter University Center for Astronomy and Astrophysics(IUCAA), Pune, India for visiting associateship. Authors thank IUCAA for local hospitality and research work as this work was done during a visit there. Authors thank Prof. Subenoy Chakraborty, Department of Mathematics, Jadavpur University for fruitful discussions.

\end{document}